\documentclass{vgtc}                          % final (conference style)
\graphicspath{{figures/}{pictures/}{images/}{./}} % where to search for the images

\usepackage{times}                     % we use Times as the main font
\usepackage{tabu}                      % only used for the table example
\usepackage{booktabs}                  % only used for the table example
\usepackage{lipsum}                    % used to generate placeholder text
\usepackage{mwe}                       % used to generate placeholder figures

\usepackage{amsmath}
\usepackage{amssymb}
\usepackage{bbding}

\vgtccategory{Research}

\vgtcinsertpkg

\title{JPEG-Inspired Cloud–Edge Holography}

\author{Shuyang Xie\textsuperscript{1,}\thanks{e-mail: sxie100@connect.hkust-gz.edu.cn\quad \textsuperscript{1}Equal contribution}\\ %
        \scriptsize HKUST(GZ) %
\and Jie Zhou\textsuperscript{1,}\thanks{e-mail: zhoujie0819@stu.scu.edu.cn}\\ %
     \scriptsize Sichuan University %
\and Jun Wang\textsuperscript{2,}\thanks{e-mail: jwang@scu.edu.cn}\\
\scriptsize Sichuan University
\and Renjing Xu\textsuperscript{2,}\thanks{e-mail: renjingxu@hkust-gz.edu.cn\quad \textsuperscript{2}Corresponding author}\\
\scriptsize HKUST(GZ)} %

\teaser{
  \centering
  \includegraphics[width=\linewidth]{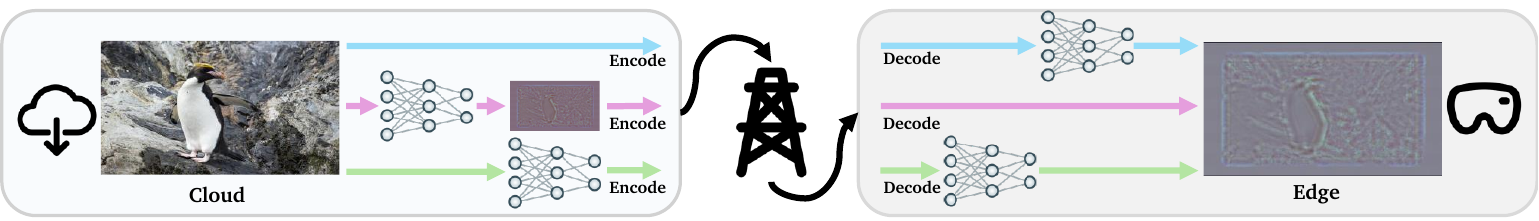}
  \caption{Cloud–to-edge strategies for computer-generated holography (CGH). Blue: transmit the target image to the edge and generate the hologram locally. Purple: offload hologram generation to the cloud, then transmit the hologram to the edge. Green: employ a neural compression pipeline with networks on both cloud and edge to reduce bandwidth usage while maintaining reconstruction quality. Proposed: building on the purple strategy, perform hologram generation entirely in the cloud, encode and transmit the hologram with a JPEG-inspired codec.}
  \label{fig1}
}

\abstract{

Computer-generated holography (CGH) presents a transformative solution for near-eye displays in augmented and virtual reality. Recent advances in deep learning have greatly improved CGH in reconstructed quality and computational efficiency. However, deploying neural CGH pipelines directly on compact, eyeglass-style devices is hindered by stringent constraints on computation and energy consumption, while cloud offloading followed by transmission with natural image codecs often distorts phase information and requires high bandwidth to maintain reconstruction quality. Neural compression methods can reduce bandwidth but impose heavy neural decoders at the edge, increasing inference latency and hardware demand. In this work, we introduce JPEG-Inspired Cloud–Edge Holography, an efficient pipeline designed around a learnable transform codec that retains the block-structured and hardware-friendly nature of JPEG. Our system shifts all heavy neural processing to the cloud, while the edge device performs only lightweight decoding without any neural inference. To further improve throughput, we implement custom CUDA kernels for entropy coding on both cloud and edge. This design achieves a peak signal-to-noise ratio of 32.15 dB at $<$ 2 bits per pixel with decode latency as low as 4.2 ms. Both numerical simulations and optical experiments confirm the high reconstruction quality of the holograms. By aligning CGH with a codec that preserves JPEG’s structural efficiency while extending it with learnable components, our framework enables low-latency, bandwidth-efficient hologram streaming on resource-constrained wearable devices—using only simple block-based decoding readily supported by modern system-on-chips, without requiring neural decoders or specialized hardware.

} % end of abstract


\begin{document}



\firstsection{Introduction}

\maketitle

% \footnotetext[1]{These authors contributed equally to this work.}
% \footnotetext[2]{Corresponding author.}

%% \section{Introduction} %for journal use above \firstsection{..} instead
% This template is for papers of VGTC-sponsored conferences which are \emph{\textbf{not}} published in a special issue of TVCG.
Computer-generated holography (CGH) is a promising engine for next-generation three-dimensional (3D) displays \cite{pi2022review, yu2023ultrahigh, Zhoulat}. By numerically simulating light diffraction, CGH produces holograms with full parallax and physically correct depth cues, improving perceptual realism and visual comfort. These advantages position CGH as a transformative solution for near-eye displays (NEDs) \cite{Zhou:25}, particularly in augmented reality and virtual reality (AR/VR) applications \cite{xiong2021augmented, wang2022development, tseng2024neural, chen2024ultrahigh}, where accurate depth perception underpins near-field hand–object manipulation, spatial alignment, and gaze-guided interaction in human–computer interaction, telepresence/teleoperation, and surgical guidance.

Among hologram encodings, phase-only holograms (POHs) are widely preferred for their high diffraction efficiency. In typical systems, POHs are rendered on spatial light modulators (SLMs) to modulate the incident wavefront and reconstruct target imagery. However, deploying CGH in compact, eyeglass-style AR/VR headsets \cite{chae2025light, np2025, nature24} is constrained by tight compute and power budgets: on-device hologram generation (\cref{fig1} blue) is both computationally and energy intensive. Cloud-assisted holographic display architectures address this limitation by offloading hologram generation to remote servers and streaming the resulting data to lightweight edge devices (\cref{fig1} purple). While offloading relieves on-device computation, it shifts the bottleneck to communication bandwidth. Conventional lossless codecs provide only modest compression for holograms, and standard lossy codecs designed for natural images introduce phase distortions that corrupt the propagated wavefront and degrade reconstruction quality. Joint methods (\cref{fig1} green) that couple hologram generation with neural compression can reduce bits per pixel (bpp) while meeting image-quality targets, but they typically require neural decoders at the edge, straining compute, memory, and real-time constraints.

This work asks whether a cloud–edge pipeline can enable lightweight, real-time hologram decoding without neural inference while preserving high-quality reconstruction at the edge. We answer in the affirmative by leveraging the ubiquitous Joint Photographic Experts Group (JPEG) standard \cite{jpeg}, which is supported across modern system-on-chip (SoC) platforms and affords extremely low decoding complexity. We introduce a JPEG-Inspired Cloud–Edge Holography pipeline that embeds parameterized transforms for compression, quantization, and reconstruction directly into the hologram generation process. These components are trained end-to-end yet leave the core JPEG computation unmodified, retaining its block-structured, hardware-friendly nature. All compute-intensive steps, hologram generation and encoding are run in the cloud; the edge device performs only lightweight decoding, eliminating neural decoders and drastically reducing on-device load. To maximize throughput, we further implement custom CUDA kernels to accelerate entropy coding/decoding for the quantized stream. Our system attains a 4.2 ms decoding latency well within real-time requirements while maintaining high-quality reconstructions at relatively low bpp.

Our main contributions are as follows:
\begin{itemize}
    \item We proposed a JPEG-Inspired Cloud–Edge Holography pipeline that removes the need for neural decoders on the device, substantially reducing edge compute, memory footprint, and latency.
    \item We designed specialized CUDA kernels for entropy encoding and decoding of the quantized data, reducing the consumption of time for encoding and decoding.
    \item We validated the effectiveness of our method for VR via simulation and optical experiments, and further demonstrated an AR system.
\end{itemize}

\section{Related work}

\subsection{Traditional Method}
Unlike conventional optical holography, CGH does not require the presence of a real object to generate the corresponding hologram, and it is not affected by external environmental factors, making the generated holograms simpler and more precise. To obtain POHs, iterative and non-iterative methods are two important approaches. Iterative methods include techniques such as Gerchberg-Saxton (GS) \cite{gs, wu}, Wirtinger Holography (WH) \cite{wh}, and stochastic gradient descent (SGD) \cite{sgd, Zhou}. GS involves continuously performing forward and backward propagation on the target image, allowing the reconstructed image after phase diffraction propagation to gradually approximate the target image. SGD treats the hologram as a learnable parameter and calculates the loss between the diffraction-propagated reconstructed image and the target image, updating the hologram to achieve convergence. Although these methods can produce high-quality holograms, they require a significant amount of time for iteration. Double phase-amplitude encoding (DPAC) \cite{ar1} is a non-iterative method that requires only a single execution to obtain a hologram, but its quality is often suboptimal.

\subsection{Learning-based Method}
Convolutional neural networks (CNNs) have garnered significant attention in the field of CGH due to their efficient data processing capabilities. A variety of networks have been developed to address challenges encountered in CGH. Peng et al. introduced HoloNet \cite{sgd}, one of the earliest approaches to use two U-Nets embedded with physical models to generate POH, taking into account optical errors such as aberrations and light source intensity. Choi et al. proposed CNNpropCNN \cite{n3d}, which uses CNNs to model physical errors at different depths, allowing for the consideration of experimental system imperfections during hologram generation. They later enhanced optical defocus effects using time-division multiplexing and also addressed different input data types \cite{tm}. Shi et al. \cite{nature} developed a lightweight network using a ResNet architecture capable of running on mobile devices to generate multi-depth holograms. Zhong et al. \cite{ccnn} proposed complex-valued convolutional neural networks, achieving breakthroughs in parameter efficiency, reconstruction quality, and inference speed. Yuan et al. \cite{compensation} further introduced a compensation network that effectively reduced ringing effects in optical experiments. Other advancements include complex-valued GANs \cite{gan}, deformable convolutions \cite{defor} and so on. For ultra-high-resolution holograms, Liu et al. \cite{liu20234k} proposed an efficient network for generating 4K resolution holograms, while Dong et al. \cite{dong2024divide} introduced the Divide-Conquer-and-Merge strategy, successfully training and inferring 8K resolution holograms on an RTX 3090 GPU for the first time.

\subsection{Compressed Holography}
Neural compression-based holographic display has emerged as a promising approach. Jiao et al. \cite{jiao2018} incorporated a neural network after the standard JPEG codec to reconstruct the lost information. Wang et al. \cite{DPRC} proposed the Dual Phase Retrieval and Compression (DPRC) framework, which significantly reduces computational cost and energy consumption. Ban et al. \cite{ban202583} introduced a learnable wave propagation model during hologram reconstruction to enhance image quality. Shi et al. \cite{shi2022neural} proposed a high-fidelity hologram compression pipeline for hologram image and video compression, while Dong et al. \cite{dong2023gaze} developed a holographic compression framework based on foveated rendering. Zhou et al. \cite{zhou2025implicit} have proposed a method called implicit feature compression, which significantly reduces the volume of original data. Recently, Qu et al. \cite{HoloZip} proposed HoloZip based on a vision transformer, achieving excellent reconstruction quality and low bpp. Despite their innovations, these methods necessitate deploying a neural decoding network on edge devices to reconstruct the hologram, which imposes a substantial memory burden on the device, and the limited computational resources, energy consumption, and inference delays further exacerbate the challenges. To enable holograms to be compatible with standard JPEG compression, Zhou et al. \cite{zhou2023end} employed an SGD-based approach to incorporate a differentiable JPEG codec into the optimization process. However, the data volume and reconstruction quality remain areas of concern, and such iterative approaches often involve repetitive computation and long execution times, making them impractical for real-time or resource-constrained display systems.

\section{Method}
\label{method}

\begin{figure}[tb]
 \centering % avoid the use of \begin{center}...\end{center} and use \centering instead (more compact)
 \includegraphics[width=\columnwidth]{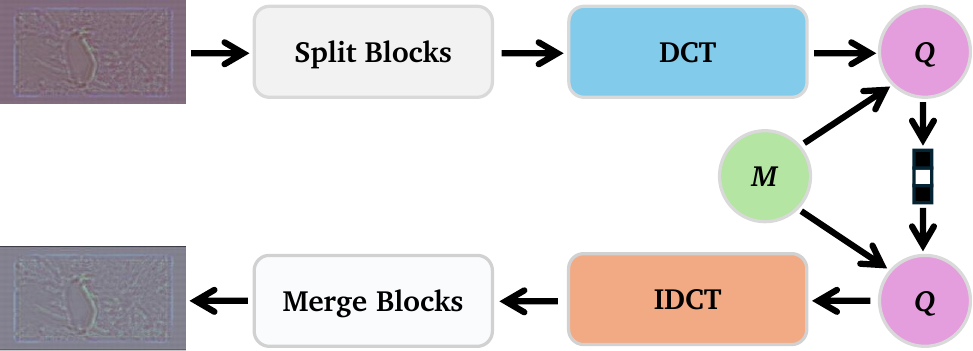}
 \caption{JPEG Flowchart. The upper section represents the encoding process: initially, the input data is split into several 8x8 blocks, followed by a DCT applied to each block, and finally, quantization is performed. The lower section illustrates the decoding process: the data stream is first dequantized, then an IDCT is applied, and finally, the blocks are merged. The quantization table \( Q \) is scaled by a learnable matrix \( M \), which is shared between both encoding and decoding processes. The parameters for DCT and IDCT are optimized independently of each other.}
 \label{fig2}
\end{figure}

\subsection{JPEG Codec}
\label{sec3.1}
Flowchart of JPEG is shown in \cref{fig2}. JPEG codec first converts the image into the YCbCr color space, separating luminance and chrominance information. The chroma channels (Cb and Cr) are typically downsampled to reduce data redundancy, here our method is gray space (only Y), we will ignore this step. Then, we split the entire image into 32,160 blocks of 8$\times$8 pixels, each 8$\times$8 block undergoes a two-dimensional discrete cosine transform (DCT), concentrating the signal energy. Finally, the DCT coefficients are quantized to achieve compression. The decoding process performs the inverse operations of the above steps in reverse order. In JPEG compression, a quantization matrix \(Q_{i,j} \in \mathbb{R}^{8 \times 8} \) is used to compute the quantized DCT coefficients $D_{i,j}$ as
\begin{equation}
D_{i,j} = \left\lfloor \frac{C_{i,j}}{ Q_{i,j}} \right\rceil
\label{eq1}
\end{equation}
where $C_{i,j}$ is the output of DCT, \( \left\lfloor \cdot \right\rceil \) denotes rounding to the nearest integer. In our approach, DCT and inverse DCT (IDCT) coefficients are learnable parameters and $Q_{i,j}$ is derived from a learnable 8$\times$8 matrix $M$ that is scaled and strictly constrained within the range from 1 to 255, after which it is quantized into integers. However, rounding operation in \cref{eq1} has a derivative of zero almost everywhere, making it incompatible with gradient-based optimization methods commonly used in deep learning. %Inspired by \cite{Lamp86}, we adopt the approximation. 
The most common approach is to use approximate methods \cite{shin2017jpeg},
\begin{equation}
\lfloor x \rceil_{\text{approx}} = \lfloor x \rceil + (x - \lfloor x \rceil)^3
\label{eq2}
\end{equation}
which provides non-zero derivatives almost everywhere and is thus more compatible with gradient-based optimization. However, this approach does not perform well in our hologram generation task. Here, we adopt the straight-through estimator to approximate the gradient of the rounding operation. During the forward pass, the input is rounded as usual. During backward propagation, however, the gradients are passed directly through the rounding function without modification—effectively treating it as an identity function. This allows the round operation to remain differentiable within the gradient-based learning framework. Let $x$ and $y$ denote the input and output, respectively. The behavior of the straight-through estimator round function can be summarized as follows:

\begin{equation}
\begin{gathered}
y = \left\lfloor x \right\rceil \\
\frac{\partial \mathcal{L}}{\partial x} = \frac{\partial \mathcal{L}}{\partial y}
\end{gathered}
\label{eq3}
\end{equation}
here, $\mathcal{L}$ is loss function. We will discuss the performance differences of the two quantization methods later.

\subsection{Entropy Coding}
Entropy coding of quantized data is a crucial step in reducing bpp. In JPEG, entropy coding typically involves arithmetic coding and Huffman coding. Due to the complexity of arithmetic coding, this paper employs Huffman coding \cite{huffman} as the method for entropy coding. To further reduce the data size, quantized data typically undergoes Zigzag scanning, differential pulse code modulation (DPCM) and run-length encoding (RLE) before Huffman coding. This process performs a Z-shaped scan on each 8$\times$8 block to produce one-dimensional data. The first number of each block represents the direct current (DC) component, while the remaining data corresponds to alternating current (AC) components. DPCM is then applied to the DC components, and RLE is used for the AC components to minimize data redundancy. Finally, entropy coding is performed on both DC and AC components using standard Huffman coding tables.

In this work, we implemented custom CUDA kernels for entropy coding, leveraging CUDA's parallelism to effectively accelerate the encoding and decoding processes. It is important to note that in DPCM, we set the prediction value to 0 instead of using the DC component of the previous block. While this choice may impact the bpp, the advantage is that it allows for multi-threading, enabling data processing without needing to wait for the previous block to be completed. Additionally, since this process is lossless, we did not include it during the training stage.

\begin{figure}[tb]
 \centering % avoid the use of \begin{center}...\end{center} and use \centering instead (more compact)
 \includegraphics[width=\columnwidth]{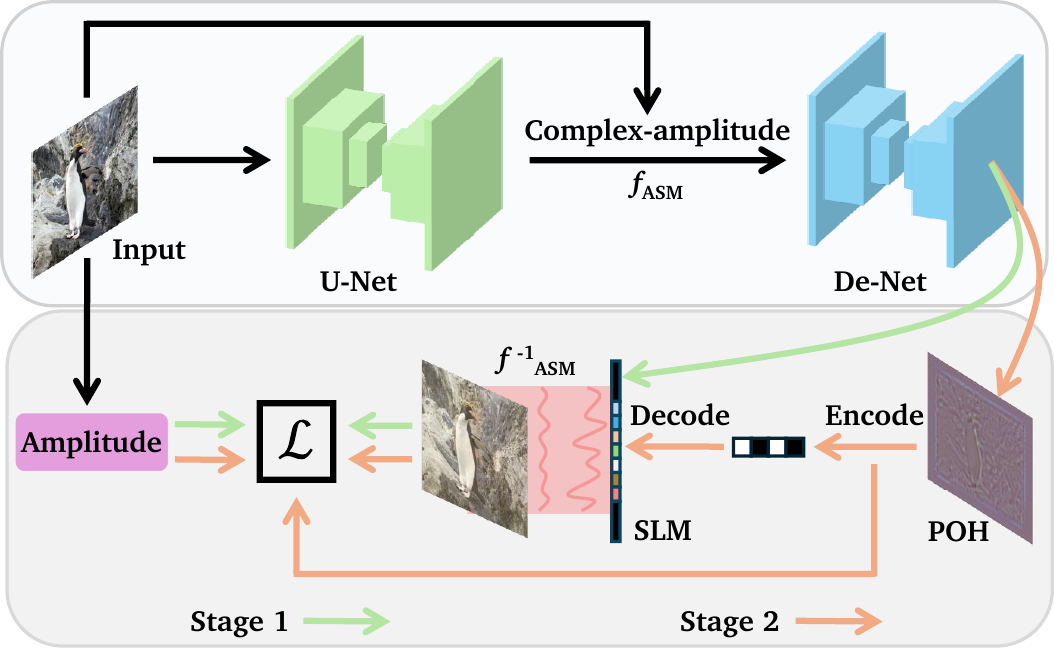}
 \caption{Pipeline of our method. The U-Net predicts the corresponding phase based on the target amplitude to form a complex-valued field at the target plane. This field is then propagated through the ASM to obtain the complex field at the SLM plane. The De-Net subsequently encodes this into a hologram. In the first training stage, the hologram is directly reconstructed through ASM without passing through a codec, and the loss is calculated based on the reconstruction. In the second training stage, the hologram is reconstructed after passing through our learnable JPEG codec, and the loss is computed against the target image, the data transformed by the DCT and the quantization table are also incorporated into the loss calculation.}
 \label{fig3}
\end{figure}

\subsection{Pipeline}
The pipeline of our proposed method is shown in \cref{fig3}. The input amplitude is first fed into a U-Net-based phase prediction network to estimate the phase. The predicted phase is then combined with the input amplitude to construct a complex field on the target plane. This complex field is propagated to the SLM plane using the angular spectrum method (ASM) \cite{asm}, resulting in the corresponding complex amplitude. The output is then passed through De-Net to generate an uncompressed hologram, which is a complex-valued deformable CNN \cite{defor}. In the first training stage, the hologram is directly reconstructed through ASM. In the second training stage, this hologram is subsequently processed by encoding and decoding, followed by another diffraction step for hologram reconstruction. The reconstructed amplitude is compared with the target amplitude to compute the loss, the data after DCT and quantization table are also the part of loss function, which is used to update the network parameters. ASM can be expressed as follow:
% \begin{small} 
\begin{equation}
\begin{gathered}
u(\phi) = \mathcal{F}^{-1}\{\mathcal{F}\{u^{i\phi}\} \mathcal{H}(f_x, f_y)\} \\
\mathcal{H}(f_x, f_y) = 
\begin{cases} 
{{\rm{e}}^{i\frac{{2\pi }}{\lambda }z\sqrt {1 - {{(\lambda {f_x})}^2} - {{(\lambda {f_y})}^2}} }}, & \text{if } \sqrt {f_x^2 + f_y^2}  < \frac{1}{\lambda } \\ 
0, & \text{otherwise}
\end{cases}
\end{gathered}
\end{equation}
% \end{small}
here, $u^{i\phi}$ denotes the optical field distribution, $\lambda$ is the wavelength, \textit{z} represents the propagation distance between the SLM plane and the target plane, $f_x$ and $f_y$ are the spatial frequencies, and $\mathcal{F}$ denotes the Fourier transform.

In our training, loss function of stage 1 is the Mean Squared Error loss function ($\mathcal{L}_{MSE}$). Compression tasks always require balancing between image quality and data stream size, a concept known as the rate-distortion tradeoff. For our compression model, the loss function of stage 2 can be expressed in terms of the input amplitude \( x \) and the output amplitude \( \hat{x} \) as follows:

\begin{equation}
\mathcal{L}(x,\hat{x}, M, C_{i,j}) = \alpha \mathcal{L}_{\text{distortion}}(x, \hat{x}) + \mathcal{L}_{\text{rate}}(M,C_{i,j})
\end{equation}
here, $M$ and $C_{i,j}$ has been mentioned in \cref{sec3.1}. $\mathcal{L}_{\text{distortion}}$ is the distortion loss and $\mathcal{L}_{\text{rate}}$ is the rate loss. 

The distortion loss is used to measure how close the reconstructed image is to the target image, and it is a decisive factor in maintaining image quality. In our case, the $\mathcal{L}_{\text{distortion}}$ can be expressed as:
\begin{equation}
\mathcal{L}_{\text{distortion}}(x, \hat{x}) = \mathcal{L}_{MSE}(x, \hat{x}) + \beta \mathcal{L}_{SSIM}(x, \hat{x})
\end{equation}
here, $\mathcal{L}_{SSIM}$ is structural similarity index (SSIM) loss function.

While maintaining image quality, it is also important to reasonably reduce the size of the data stream. The rate loss plays a crucial role in this, as its configuration will determine the final bpp. In our case, the $\mathcal{L}_{\text{rate}}$ can be expressed as:
\begin{align}
% \begin{equation}
\mathcal{L}_{\text{rate}}(M,C_{i,j}) &= \underset{\mathcal{L}_1}{\underbrace{\|1/M\|_1}}
\notag
\\&+ \underset{\mathcal{L}_2}{\underbrace{\gamma |\text{CDF}(C_{i,j}+0.5)-\text{CDF}(C_{i,j}-0.5)|}}
% \end{equation}
\end{align}
The first part of rate loss involves calculating the L1 norm of the reciprocal of $M$ \cite{strumpler2020learning}. CDF is cumulative distribution function, the second part of rate loss involves adding noise to $C_{i,j}$ and then calculating the cumulative probability to obtain likelihood values \cite{balle2018variational}. $\alpha$, $\beta$ and $\gamma$ are loss weights. We will discuss the roles of different loss function components in detail later.

\section{Model train and experiment}
\label{sec4}

\subsection{Training Details}
To assess the effectiveness of the proposed method, we compared it against three representative baselines: SGD with the standard torchvision JPEG codec (SGD+JPEG), HoloNet+JPEG, and DPRC. All implementations were developed in Python 3.9 using the PyTorch 2.1.1 framework and executed on a Linux workstation equipped with an AMD EPYC 7543 CPU and an NVIDIA GeForce RTX 3090 GPU. The parameters were set as follows: the pixel pitch of SLM was 8 $\mu$m, and the propagation distance was fixed at 20 cm. The input image resolution was set to 1600$\times$880, while the hologram resolution was set to 1920$\times$1072. Values of $\beta$ and $\gamma$ were set to 0.007 and 0.001, respectively, $\alpha \in \{0.7, 0.8, 1.0\}$ corresponds to different levels of the rate-distortion tradeoff. 
For training, our models were optimized for 20 epochs at each stage with a batch size of 1 and an initial learning rate of 0.001, using the DIV2K training dataset \cite{div2k}. To ensure fairness in comparison, DPRC was trained for 20 epochs per stage, HoloNet for a total of 40 epochs, and the SGD baseline was optimized for 1000 iterations. Performance was evaluated on the DIV2K validation dataset. Quantitative comparisons were conducted using PSNR and bpp to jointly measure reconstruction quality and compression efficiency.
\subsection{Simulation Results}

The standard Huffman table is a set of codes developed by researchers based on statistical characteristics of natural images, which offers relatively good performance in general scenarios. However, its compression efficiency is suboptimal for holograms. Therefore, in addition to employing the standard table for lossless compression, we introduced an optimization approach: generating a customized Huffman table tailored to the content of the hologram. In essence, this approach involves counting the frequency of each symbol in the image, assigning shorter codewords to those that occur more frequently, and longer codewords to those that appear less often, this method can further reduce the bpp. Nevertheless, since generating such a table is computationally intensive and not well-suited for CUDA-based parallel processing, the Huffman encoding and decoding in the optimized method were implemented without CUDA acceleration, leading to an increase in decoding time.

As shown in \Cref{speed}, DPRC requires the longest decoding time, reaching 52.2 ms, while our method takes only 4.2 ms, which is significantly faster than DPRC. When using the optimized Huffman table for decoding, although the time increases compared to the baseline method, it remains faster than DPRC. This demonstrates the superior decoding efficiency of our approach on edge devices, comfortably meeting the requirements for real-time decoding. Note that when measuring the decoding time for all methods, it is essential to use the command \textit{torch.cuda.synchronize()} to synchronize the CPU and GPU operations; otherwise, the computed average time may be inaccurate.

\begin{figure}[tb]
 \centering % avoid the use of \begin{center}...\end{center} and use \centering instead (more compact)
 \includegraphics[width=\columnwidth]{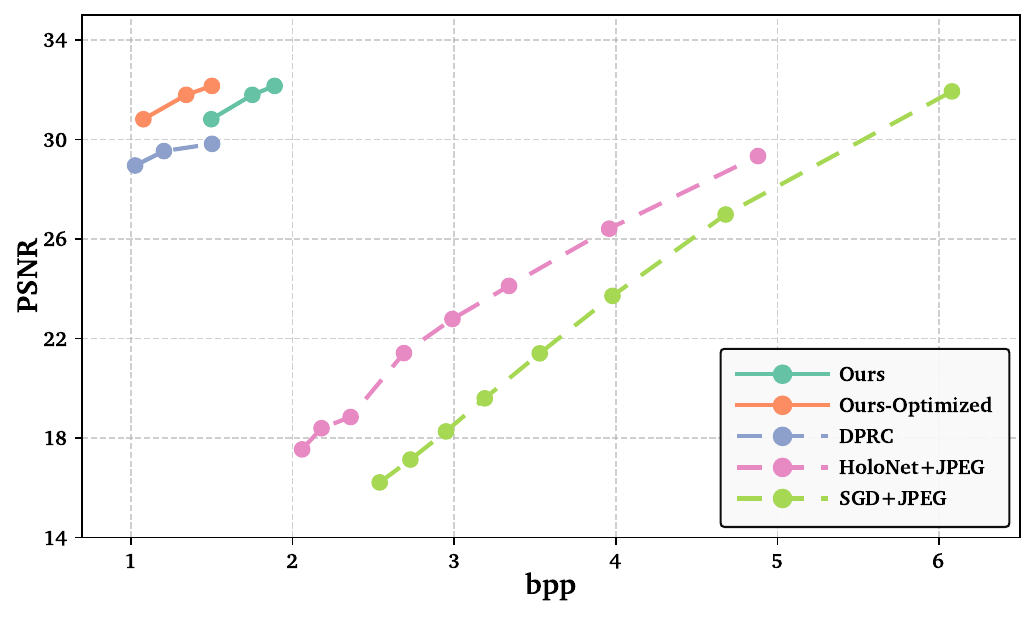}
 \caption{Quantitative evaluation results of different methods. The bpp calculation for our optimized approach includes the overhead of the generated Huffman table. Under similar bpp conditions, our method achieves the highest PSNR.}
 \label{fig4}
\end{figure}

Quantitative evaluation results of different methods are illustrated in \cref{fig4}. It can be observed that although both SGD and HoloNet can achieve reconstruction quality comparable to our method, they require a higher bpp to attain such performance. When their bpp is reduced to a level similar to ours, their reconstruction quality becomes significantly worse than that of our approach. Compared to DPRC, our optimized method achieves higher PSNR under similar bpp conditions. Although the bpp of our method using the standard encoding table is slightly higher than that of DPRC, our decoding speed is considerably faster.

\begin{table}[tb]
    \setlength{\tabcolsep}{1pt}
  \caption{Decoding time for different compression methods}
  \scriptsize%
	\centering%
  \begin{tabu}{%
   X[c]% 第一列居中
    *{3}{X[c]}% 剩余4列都居中
	}
\toprule
Method & DPRC &  Ours & Ours-Optimized \\
\midrule
Dec. time (ms) & 52.2 & 4.2  & 14.9 \\
\bottomrule
\end{tabu}%
\label{speed}
\end{table}

\begin{figure*}[tb]
 \centering % avoid the use of \begin{center}...\end{center} and use \centering instead (more compact)
 \includegraphics[width=\textwidth]{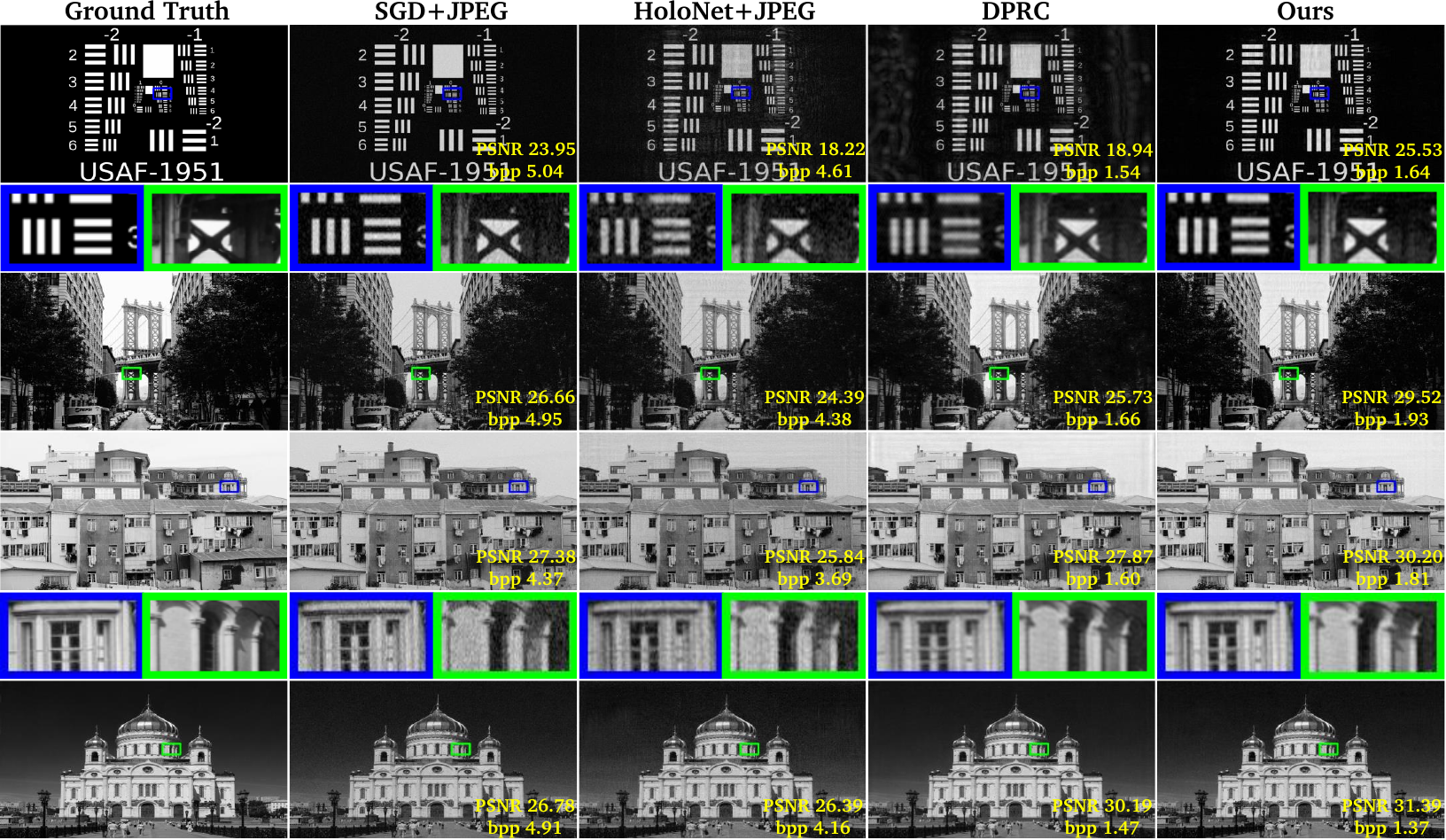}
 \caption{Several representative examples from the simulation experiments. Quality level of JPEG codec in SGD and HoloNet was set to 90 here.}
 \label{fig5}
\end{figure*}

Several examples from the simulation experiments are presented in \cref{fig5}. When encoding and decoding both SGD and HoloNet using the JPEG codec from torchvision, the quality level was set to 90 in this figure. It can be clearly observed that, compared to our method, both SGD and HoloNet exhibit significant noise. Although DPRC achieves a lower bpp, our method delivers superior contrast and higher PSNR without significantly increasing the bpp.

\subsection{Optical Results}

\begin{figure}[tb]
 \centering % avoid the use of \begin{center}...\end{center} and use \centering instead (more compact)
 \includegraphics[width=\columnwidth]{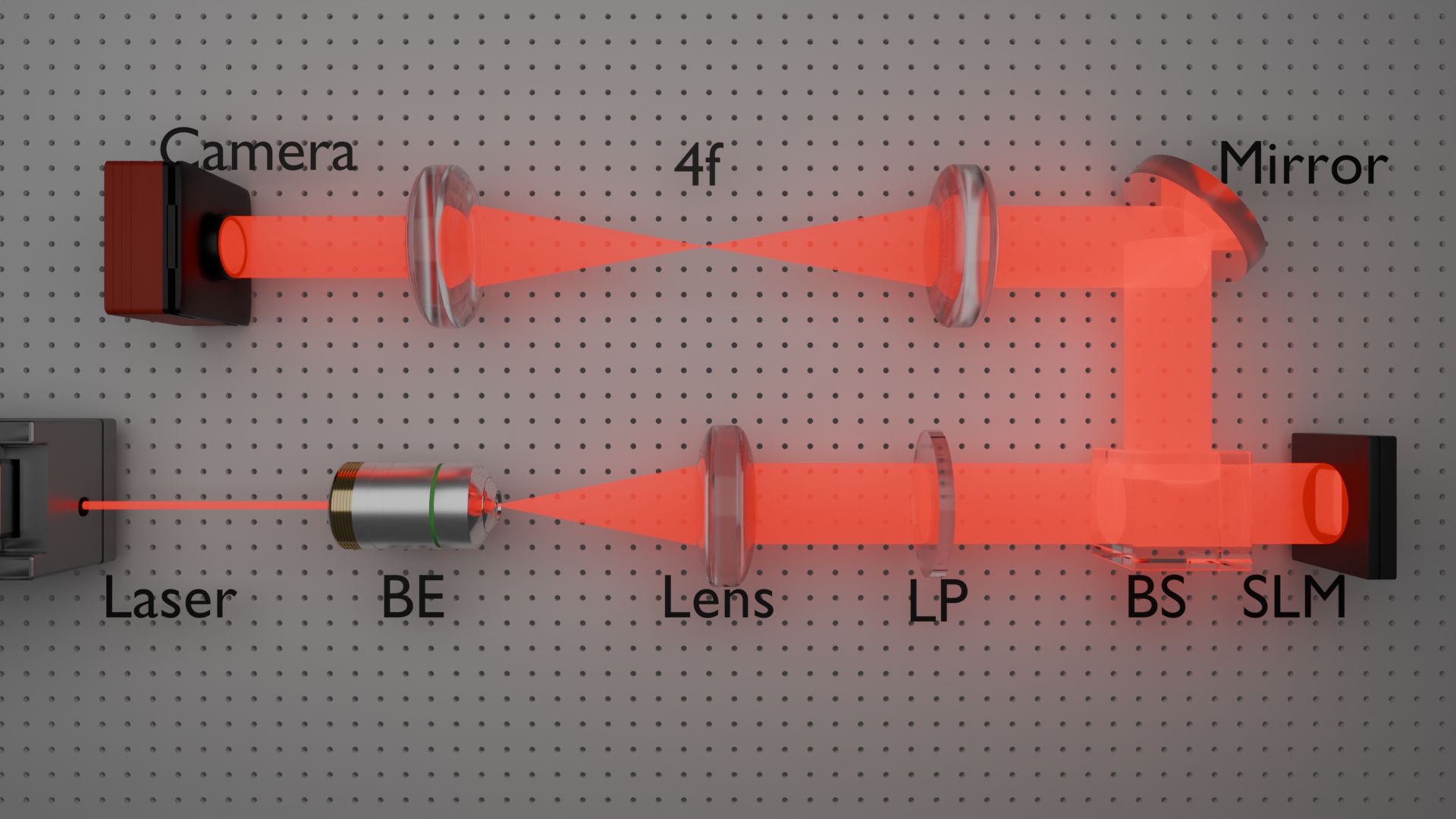}
 \caption{Schematic diagram of the optical experimental setup. BE: Beam Expander; LP: Linear Polarizer; BS: Beam Splitter; SLM: Spatial Light Modulator.}
 \label{fig6}
\end{figure}

The optical experimental setup employed in this study is shown in \cref{fig6}. The laser beam first passes through a beam expander (BE) and is subsequently collimated by a lens. After transmission through a linear polarizer (LP), the beam illuminates the spatial light modulator (SLM). The modulated light is then reflected and directed to a beam splitter (BS), where a portion of the beam is transmitted into the detection path and subsequently redirected by a mirror. Finally, the beam passes through a 4f filtering system, which removes higher-order diffraction components, before being recorded by the camera.

\begin{figure*}[tb]
 \centering % avoid the use of \begin{center}...\end{center} and use \centering instead (more compact)
 \includegraphics[width=\textwidth]{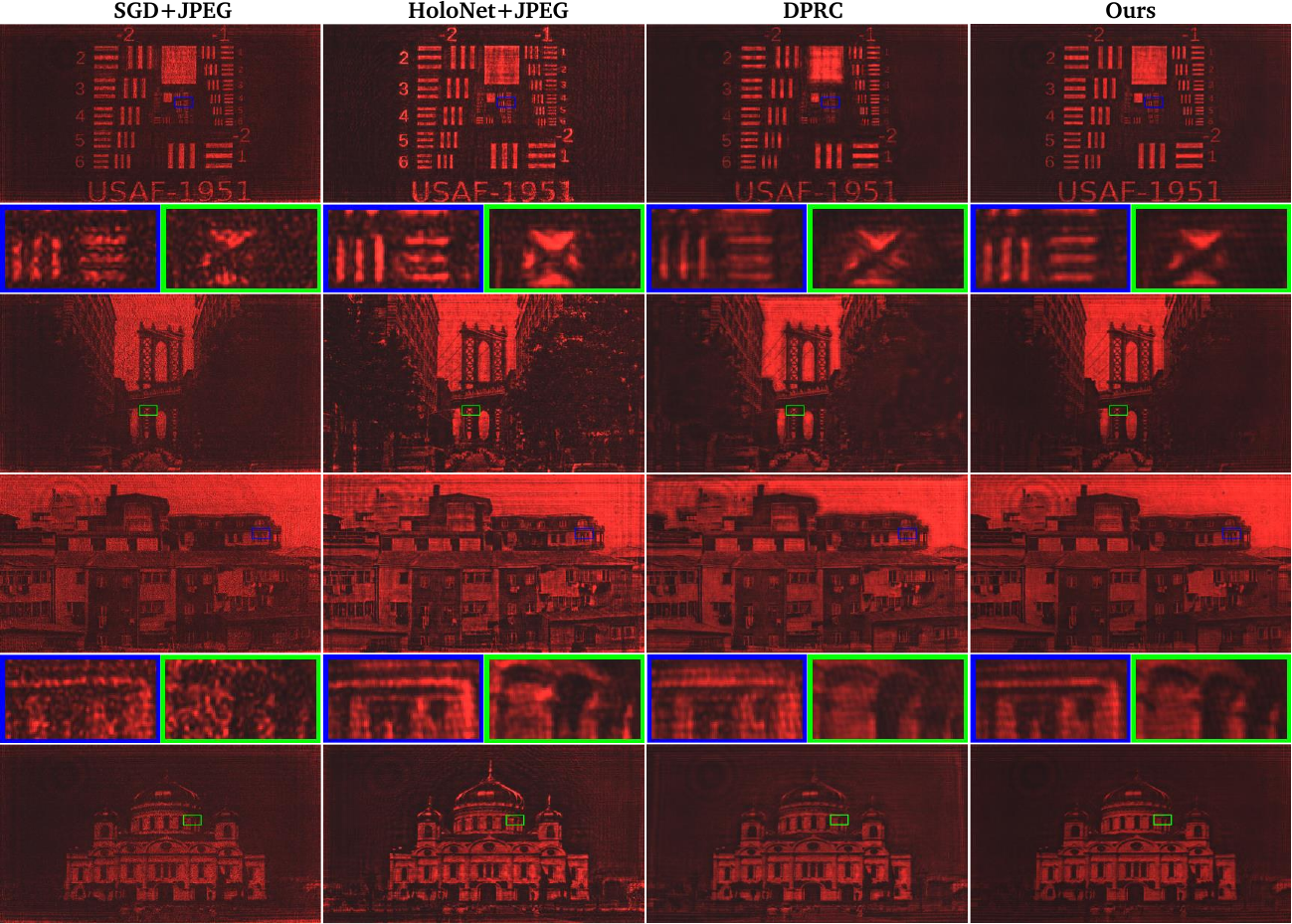}
 \caption{Comparative optical VR experiment. SGD+JPEG exhibits pronounced speckle noise, HoloNet+JPEG introduces optical artifacts, and DPRC suffers from blurred details, whereas the proposed method effectively suppresses speckle noise and artifacts while preserving fine structural features, achieving superior reconstruction quality.}
 \label{fig7}
\end{figure*}

The results of a comparative optical VR experiment are presented in \cref{fig7}. The SGD+JPEG baseline exhibits pronounced speckle noise, consistent with the simulation results. HoloNet+JPEG reduces speckle noise to some extent but introduces noticeable optical artifacts, particularly in darker regions. DPRC provides relatively higher overall quality, however, its reconstructions suffer from blurred high-frequency details. In contrast, the proposed method achieves superior reconstruction quality, effectively suppressing speckle noise and artifacts while preserving fine structural features.

\begin{figure}[tb]
 \centering % avoid the use of \begin{center}...\end{center} and use \centering instead (more compact)
 \includegraphics[width=\columnwidth]{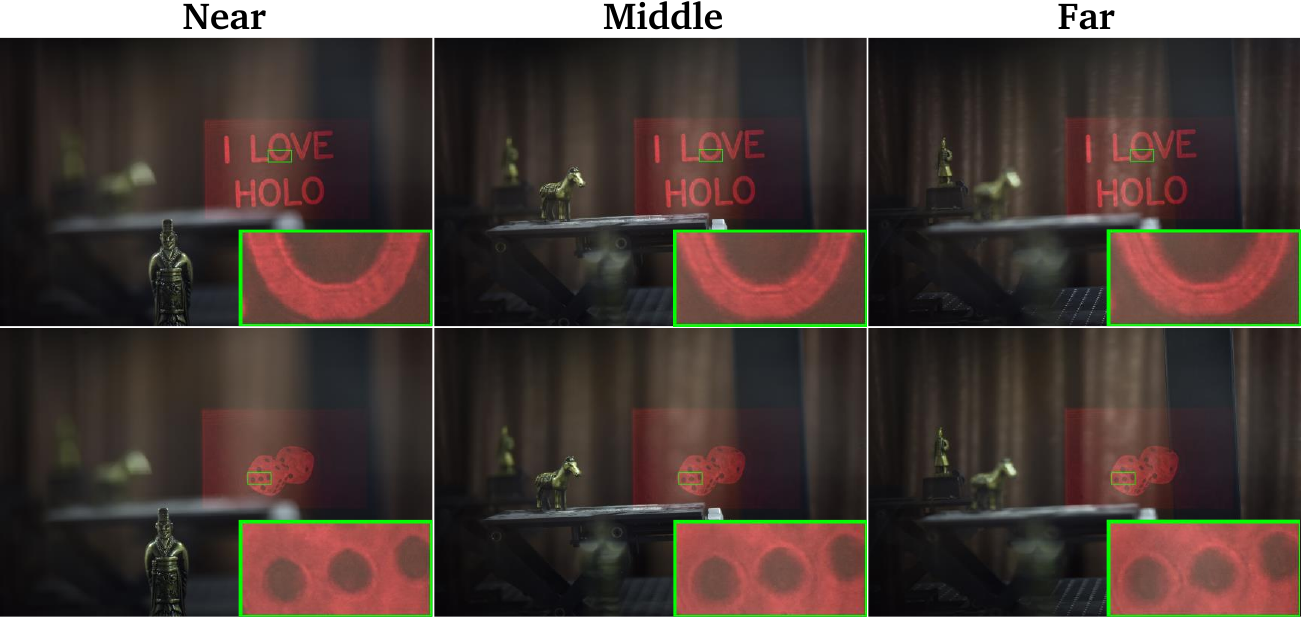}
 \caption{Optical AR experiment. The observations indicate that both physical and virtual objects exhibit depth-dependent focus and defocus characteristics.}
 \label{fig8}
\end{figure}

To further assess the effectiveness of the proposed method, we conducted an AR experiment, as illustrated in \cref{fig8}. By varying the camera parameters, optical results were captured at three distinct depth planes. The observations indicate that both physical and virtual objects exhibit depth-dependent focus and defocus characteristics, thereby providing strong evidence that the proposed method faithfully reproduces depth cues and enhances the perceptual realism of holographic displays.

\section{Ablation study}

\subsection{Rounding Operation}
\begin{figure}[tb]
 \centering % avoid the use of \begin{center}...\end{center} and use \centering instead (more compact)
 \includegraphics[width=\columnwidth]{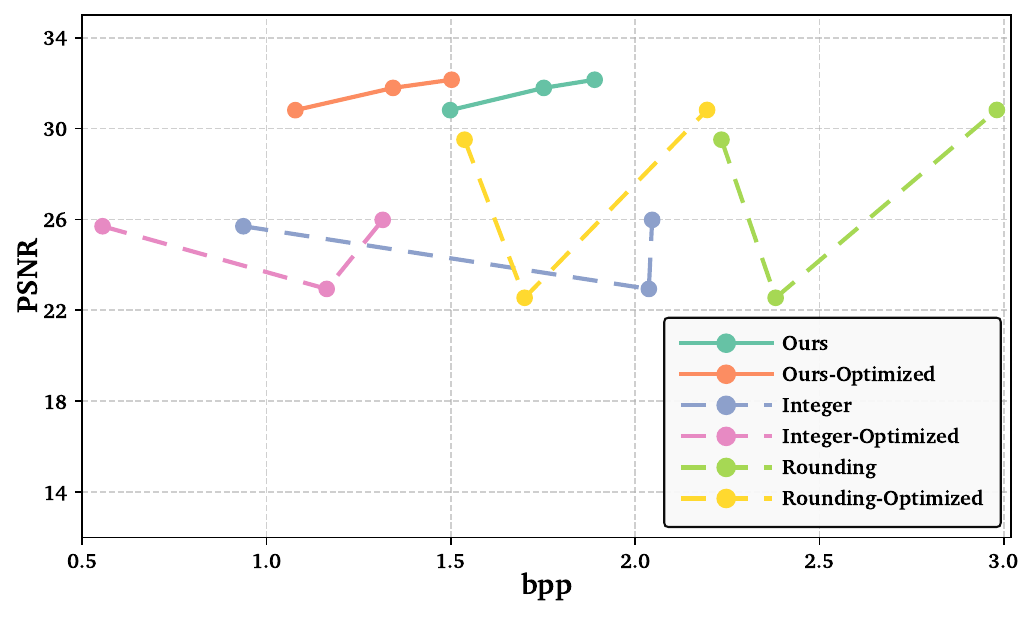}
 \caption{Quantitative evaluation of different approximate gradient methods. The terms "integer" and "rounding" refer to two methods capable of lossless compression during inference for \cref{eq2}. "Optimized" denotes the approach where a customized Huffman table is generated adaptively based on the content of each hologram.}
 \label{fig9}
\end{figure}

In this subsection, we will discuss the performance differences when using \cref{eq2} and \cref{eq3}. It is important to note that during entropy coding with a standard Huffman table, the input data must be in integer format. However, applying the rounding operation to the quantized data using \cref{eq2} inevitably introduces floating-point numbers. The primary reason for not directly employing optimized Huffman tables for floating-point data is that, within a given range, such values tend to exhibit significantly greater variability compared to integers. Generating dedicated Huffman tables for floating-point data is computationally expensive and often fails to effectively reduce data redundancy. To address this, we employ two methods to convert the data into integers for comparison with our approach. The first method involves directly casting the data type to an integer (int8 or int16), while the second method applies an additional rounding operation to \cref{eq2} during the inference process. The fundamental difference between these two approaches lies in truncation versus rounding, which can lead to significant performance variations. The results are presented in \cref{fig9}. After incorporating the \cref{eq2} method to make our approach differentiable, we observed that, regardless of the encoding technique applied to the data stream during inference, the reconstructed images exhibit either a lower PSNR or a higher bpp. This outcome confirms the effectiveness of the straight-through estimator round function employed in our method.

\subsection{Learnable Parameter}
In this subsection, we evaluate the effectiveness of incorporating learnable parameters into the JPEG codec. We will conduct eight comparative experiments in which the key variable is the learnability of the parameters in the JPEG codec. Experimental results are presented in \Cref{learn}.

When non-learnable parameters are used in the DCT and IDCT processes, the matrix $M$ plays a critical role in determining the reconstruction quality. This is primarily due to the scaling factor being consistently set to 100 across all our training configurations, a significant portion of high-frequency information will be omitted, which considerably compromises reconstruction quality. When $M$ is introduced as a learnable parameter, the reconstruction quality improves significantly, albeit at the cost of an increase in bpp. By making both the DCT and IDCT learnable, further improvements in reconstruction quality are achieved while keeping the bpp at a comparable level. When all parameters are made learnable, the optimal rate-distortion trade-off is attained.

\begin{table}[tb]
  \caption{Learnability of Different Components in JPEG}
  % \label{tab:vis_papers}
  \scriptsize%
	\centering%
  \begin{tabu}{%
   X[c]% 第一列居中
    *{4}{X[c]}% 剩余4列都居中
	}
\toprule
DCT & IDCT & $M$ & PSNR & bpp  \\
\midrule
\XSolidBrush & \XSolidBrush & \XSolidBrush & 20.99 & 0.93 \\
\XSolidBrush & \XSolidBrush & \CheckmarkBold  & 29.16 & 1.61 \\
\XSolidBrush & \CheckmarkBold & \XSolidBrush  & 23.34 & 1.54 \\
\XSolidBrush & \CheckmarkBold & \CheckmarkBold  & 29.98 & 1.60 \\
\CheckmarkBold & \XSolidBrush & \XSolidBrush & 20.95 & 0.92  \\
\CheckmarkBold & \XSolidBrush & \CheckmarkBold & 28.86 & 1.56  \\
\CheckmarkBold & \CheckmarkBold & \XSolidBrush & 31.73 & 1.60  \\
  % \midrule
\CheckmarkBold & \CheckmarkBold & \CheckmarkBold & 32.15 & 1.50  \\
\bottomrule
\end{tabu}%
\label{learn}
\end{table}

\subsection{Loss Function}
In this subsection, to demonstrate the effectiveness of our rate loss, we will primarily discuss the role of each component within the rate loss. We will also present four corresponding experiments. Here, we will use $\mathcal{L}_{\text{1}}$ and $\mathcal{L}_{\text{2}}$ to represent the two components of the $\mathcal{L}_{\text{rate}}$. Our results are shown in \Cref{losst}.

When neither $\mathcal{L}_{\text{1}}$ nor $\mathcal{L}_{\text{2}}$ is included, the model achieves the best reconstruction quality, albeit at the highest bpp. Introducing either $\mathcal{L}_{\text{1}}$ or $\mathcal{L}_{\text{2}}$ individually leads to a reduction in bpp, while incorporating both yields the lowest bpp. This reduction in bpp, however, is generally accompanied by a decrease in reconstruction quality. The primary aim of this experiment is to verify the effectiveness of the proposed loss function in reducing bpp rather than keeping reconstruction image quality.

\begin{table}[tb]
  \caption{Impact of Different Components in the Rate Loss on bpp Reduction}
  \scriptsize%
	\centering%
  \begin{tabu}{%
   X[c]% 第一列居中
    *{4}{X[c]}% 剩余4列都居中
	}
\toprule
$\mathcal{L}_{\text{distortion}}$ & $+\mathcal{L}_{\text{1}}$ &  $+\mathcal{L}_{\text{2}}$ & PSNR & bpp  \\
\midrule
\CheckmarkBold & \XSolidBrush & \XSolidBrush  & 37.33 & 6.39 \\
\CheckmarkBold & \CheckmarkBold & \XSolidBrush & 35.88 & 4.73  \\
\CheckmarkBold & \XSolidBrush & \CheckmarkBold & 32.77 & 2.00 \\
  % \midrule
\CheckmarkBold & \CheckmarkBold & \CheckmarkBold & 32.15 & 1.50  \\
\bottomrule
\end{tabu}%
\label{losst}
\end{table}

\section{Limitation and Future work}

% The optimized Huffman tables are generated solely on the cloud side and transmitted along with the data stream to the edge devices for decoding. As such, the optimized approach is also amenable to parallelization to achieve shorter decoding times, which we plan to explore in future work. 
Holographic video display involves sequentially loading POHs of video frames onto a SLM for reconstruction. Although this work does not attempt to compress video frames, the exploration of efficient video compression schemes constitutes an exciting direction for future research. Furthermore, due to hardware limitations, our optical experiments were conducted only with a monochromatic laser source without color reproduction. In addition, the current holographic representations are limited to a single plane rather than multi-depth scenes. We intend to integrate the proposed method into more advanced frameworks in the future to further investigate its applicability under broader conditions.

\section{Conclusion}
In this paper, we propose an efficient CGH pipeline centered around a learnable transform codec that retains the block-structured and hardware-friendly nature of JPEG. To further boost decoding performance, we implemented custom CUDA kernels that enable real-time decoding speed. To the best of our knowledge, it is the first hologram compression model capable of efficient inference on edge devices without relying on neural networks. Trained with our tailored strategy, the method demonstrates consistent effectiveness and robustness in both simulations and optical experiments. Our work paves the way for next-generation AR/VR applications, offering the potential to significantly reduce power consumption and latency, thereby enabling lighter, more responsive, and more immersive user experiences.
% \lipsum[1]%

% \section*{Supplemental Materials}
% \label{sec:supplemental_materials}

% Refer to the instructions for this section (\cref{sec:supplement_inst}).
% Below is an example you can follow that includes the actual supplemental material for this template:

% All supplemental materials are available on OSF at \url{https://doi.org/10.17605/OSF.IO/2NBSG}, released under a CC BY 4.0 license.
% In particular, they include (1) Excel files containing the data for and analyses for creating \cref{tab:vis_papers} and \cref{fig:vis_papers}, (2) figure images in multiple formats, and (3) a full version of this paper with all appendices.
% Our other code is intellectual property of a corporation---Starbucks Research---and there is no feasible way to share it publicly.

% \section*{Figure Credits}
% \label{sec:figure_credits}

% Refer to the instructions for this section (\cref{sec:figure_credits_inst}).
% Here are the actual figure credits for this template:

% \Cref{fig:teaser} image credit: Scott Miller / Special to the Vancouver Sun, January 22, 2009, page A6.

% \Cref{fig:vis_papers} is a partial recreation of Fig.\ 1 from \cite{Isenberg:2017:VMC}, which is in the public domain.

% %% if specified like this the section will be committed in review mode
% The authors wish to thank A, B, and C. This work was supported in part by
% a grant from XYZ.}

%\bibliographystyle{abbrv}
\bibliographystyle{abbrv-doi}

\bibliography{template}
\end{document}